\documentclass{ws-ijmpe}
\usepackage[super,compress]{cite}
\usepackage{soul}
\usepackage{xcolor}
\begin{document}
	
	\markboth{Usuf Rahaman }{}
	
	\catchline{}{}{}{}{}
	
	\title{Investigation of ground state properties and shape evolution in Hf isotopes using the CDFT approach}
	
	\author{Usuf Rahaman\footnote{Email: urahaman@myamu.ac.in}}
	
	\address{Department of Physics, Madanapalle Institute of Technology $\&$ Science, Madanapalle-517325, India.}

	\maketitle
	
	\begin{history}
		\received{Day Month Year}
		\revised{Day Month Year}
	\end{history}
	
	\begin{abstract}
The ground-state properties and shape evolution of even-even hafnium isotopes ranging from $N=80$ to the neutron dripline are thoroughly examined using Covariant Density Functional Theory (CDFT) with density-dependent effective interactions, specifically the parameter sets DD-ME1, DD-ME2, DD-PC1, and DD-PCX. 
Key nuclear properties, including binding energies, two-neutron separation energies ($S_{2n}$), two-neutron shell gaps ($\delta S_{2n}$), neutron pairing energies ($E_{pair,n}$), quadrupole deformation parameters ($\beta_2$), root-mean-square (RMS) charge and matter radii, and neutron skin thickness ($\Delta r_{np}$), are systematically computed and compared with available experimental results and predictions from various theoretical models. These include the Hartree-Fock-Bogoliubov (HFB) framework employing the Skyrme SLy4 interaction, the Finite Range Droplet Model (FRDM),  the deformed relativistic Hartree–Bogoliubov theory in continuum (DRHBc) using the PC-PK1 functional, and the relativistic mean-field (RMF) approach with NL3 parameterization.
Shell closures at $N = 82$ and $N = 126$, subshell effects at $N = 108$ and $N = 152$, and shape transitions with coexistence in $^{192}$Hf and $^{222-236}$Hf are observed. Neutron skin thickness increases with neutron excess, and potential energy surfaces show consistent trends, validating CDFT’s reliability for nuclear structure predictions.
	\end{abstract}
	
	\keywords{ Covariant Density Functional Theory, Magic Number, Neutron Skin thickness, Neutron-rich nuclei.}
	
	\ccode{PACS numbers:}
	
	
	\section{Introduction}
	The study of transitional nuclei is a cornerstone in the field of nuclear physics, offering profound insights into the structural properties of atomic nuclei, particularly their prolate and oblate configurations~\cite{Hamamoto2009, Hamamoto2012, Dennis2024, kiss2024, stev}. These nuclei, characterized by dynamic and static effects, have been the subject of extensive theoretical and experimental investigations.  In recent decades, a good number of experiments and theoretical models, both relativistic and non-relativistic, have been applied to explore transitional nuclei within the atomic number range \( \mathrm{Z} = 72-80 \)
	~\cite{alkh, whel,Rahaman:2024wdi, Rahaman:2020zqn, Usmani:2017pgj, john, pt1, pt2, Parveen2016, USQCD}. These studies have significantly advanced our understanding of axial and triaxial shapes and their evolution in isotopic chains, particularly in the context of neutron-rich nuclei ~\cite{rob2009,rod2010, Sarriguren2008, Nomura1, Nomura2, Nomura3, Ramos2014}.
	
	One of the most intriguing phenomena observed in this region is the coexistence of superdeformed shapes, which disrupt the inherent spherical symmetry of nuclei and introduce non-sphericity governed primarily by the quadrupole deformation~\cite{Robert1991}.Remarkably, some nuclei exhibit multiple deformations at nearly identical excitation energies, a phenomenon known as shape coexistence ~\cite{morinaga1956, bonatsos2023}.  This opens the possibility of studying quantum shape oscillations or mixing between different configuration, making transitional nuclei highly promising candidates for structural studies ~\cite{john}. 
	 The mass region around  $A \sim 190$ is particularly fascinating, as it represents a critical transition zone between prolate and oblate shapes, which has been thoroughly investigated in recent years ~\cite{sharma2025, dadwal2017, uma2015, khalaf2016, jain2023, sharma2013, djerroud2000, barea2009}.
	
	The isotopes of hafnium (Hf) play a pivotal role in the transitional region, where nuclear shapes evolve from spherical to prolate or oblate configurations and eventually return to spherical shapes as the neutron number increases~\cite{liu2011, hartley2005, bochev1977, mccutchan2008, chowdhury1999, vasileiou2023}. This region provides a unique framework for examining shape evolution, particularly in the context of neutron-rich nuclei. Our study focuses on the even-even hafnium isotopes, which span from the beta-stable region to the neutron-drip line. These nuclei exhibit intriguing phenomena such as neutron halos and skins, which are strongly influenced by pairing correlations ~\cite{tanihata1996,horo2001,Horowitz2001}. Consequently, an explicit treatment of pairing correlations is essential for accurately describing nuclei near the drip line.

	Density Functional Theory (DFT) has emerged as one of the most powerful and versatile approaches for describing nuclear properties across the entire chart of nuclides. Among the various implementations of nuclear DFT, covariant density functional theory (CDFT)~\cite{niksic2014,lalazissis2005,roca2011,niksic2002a,agbemava2014} has demonstrated particular success through its use of Lorentz-covariant energy density functionals. The CDFT framework offers several fundamental advantages that make it especially suitable for nuclear structure studies. First, it provides a natural description of spin-orbit interactions~\cite{ring1996, meng2006}, which are crucial for explaining nuclear shell structure, without requiring phenomenological adjustments. Second, it offers a clear relativistic interpretation of pseudospin symmetry~\cite{liang2015} through its treatment of scalar and vector potentials, explaining the observed near-degeneracy of certain orbitals. The theoretical foundation of CDFT is further strengthened by its connection to relativistic Brueckner-Hartree-Fock theory~\cite{shen2019}, which links the approach to realistic nucleon-nucleon interactions. Additionally, the covariant framework consistently incorporates time-odd mean fields~\cite{zhao2018, zhang2014} that are essential for describing rotating nuclei, odd-mass systems, and magnetic properties. Valuable insights have also been gained through the nonrelativistic reduction of covariant functionals~\cite{ren2020}, which helps bridge the connection between relativistic and nonrelativistic approaches. These theoretical advantages, comprehensively discussed in Ref.~\cite{RelDFTBook}, make CDFT particularly well-suited for investigating ground-state properties and shape evolution across the nuclear landscape~\cite{abusara2012,agbemava2015,agbemava2017}, while providing fundamental insights into diverse nuclear structure phenomena~\cite{meng2015,matev2007,afanasjev2008,afana2016}.

	In this research, we utilize Covariant Density Functional Theory (CDFT) to examine the transition from spherical to prolate or oblate shapes in the even-even hafnium isotopic chain, covering masses from the beta-stable region to the neutron-drip line. This mass range provides an intriguing opportunity to study neutron magic numbers and localized regions of shape phase transitions, which may be energetically favorable for nuclear reaction processes. Our calculations yield insights into key nuclear characteristics, such as ground-state energy, quadrupole deformation, nuclear radii, skin thickness, and potential energy surfaces. Although triaxial symmetry is significant in these nuclei, this study concentrates solely on axially deformed configurations.

	This investigation not only enhances our understanding of nuclear shape transitions but also contributes to the broader field of nuclear physics by providing a systematic and comprehensive analysis of hafnium isotopes across a wide range of neutron numbers. The insights gained from this study have significant implications for nuclear structure theory, astrophysics, and the synthesis of heavy elements in stellar environments ~\cite{doba1984,chabanat1998, erler2012}.

	The structure of this paper is outlined as follows. In Section 2, we present a detailed description of the theoretical framework and the computational methods used in this study. Section 3 is dedicated to the presentation and discussion of the results obtained. Finally, Section 4 offers a summary of the main conclusions drawn from the analysis.

	\section{Theoretical Framework}
	
	\subsection{Covariant Density Functional Theory}
	
	In this study, we utilize two distinct approaches within the framework of covariant density functional theory (CDFT) ~\cite{niksic2014, niksic2002a, cescato1998}: the density-dependent meson-exchange (DD-ME) model ~\cite{lalazissis2005, niksic2002b} and the density-dependent point-coupling (DD-PC) model ~\cite{niksic2008, yuksel2019}. These models differ fundamentally in their handling of nuclear interactions. The DD-ME model employs a finite-range interaction mediated by meson exchanges, whereas the DD-PC model adopts a zero-range contact interaction, augmented by a gradient term in the scalar-isoscalar sector. By leveraging these complementary formulations, we aim to investigate the influence of interaction range on nuclear properties and the predictive power of our theoretical results.
	
	\subsubsection{The Meson-Exchange Model}
	
	The meson-exchange model conceptualizes the nucleus as a collection of Dirac nucleons interacting via the exchange of mesons with finite masses, leading to interactions with a non-zero range~~\cite{niksic2014,lalazissis2005,niksic2002b, ring1996, vretenar2005, typel1999, brockmann1992, hofmann2001}. The key meson fields in this framework include the isoscalar-scalar $\sigma$ meson, the isoscalar-vector $\omega$ meson, and the isovector-vector $\rho$ meson, which together provide a robust description of nuclear structure across the isotopic chart. This model is grounded in a Lagrangian density~~\cite{gambhir1990, serot1986, serot1997} that incorporates density-dependent interaction vertices:
	
	\begin{eqnarray}
		{\cal L} &=& \bar{\psi} \left[ \gamma (i\partial - g_{\omega}\omega - g_{\rho} \vec{\rho} \, \vec{\tau} - eA) - m - g_{\sigma}\sigma \right] \psi \nonumber \\
		&+& \frac{1}{2} (\partial \sigma)^2 - \frac{1}{2} m_{\sigma}^2 \sigma^2 - \frac{1}{4} \Omega_{\mu\nu} \Omega^{\mu\nu} + \frac{1}{2} m_{\omega}^2 \omega^2 \nonumber \\
		&-& \frac{1}{4} \vec{R}_{\mu\nu} \vec{R}^{\mu\nu} + \frac{1}{2} m_{\rho}^2 \vec{\rho}^{\,2} - \frac{1}{4} F_{\mu\nu} F^{\mu\nu}.
		\label{lagrangian_me}
	\end{eqnarray}
	
	Here, $\psi$ denotes the Dirac spinor field of the nucleons, $m$ is the nucleon mass, and $e$ represents the proton charge (zero for neutrons). The meson masses are denoted by $m_{\sigma}$, $m_{\omega}$, and $m_{\rho}$, with corresponding coupling strengths $g_{\sigma}$, $g_{\omega}$, and $g_{\rho}$. These couplings, along with the meson masses, are adjustable parameters of the model. The field tensors $\Omega^{\mu\nu}$, $\vec{R}^{\mu\nu}$, and $F^{\mu\nu}$ describe the dynamics of the $\omega$, $\rho$, and electromagnetic fields, respectively, and are defined as:
	
	\begin{equation}
		\Omega^{\mu\nu} = \partial^{\mu} \omega^{\nu} - \partial^{\nu} \omega^{\mu},
	\end{equation}
	\begin{equation}
		\vec{R}^{\mu\nu} = \partial^{\mu} \vec{\rho}^{\nu} - \partial^{\nu} \vec{\rho}^{\mu},
	\end{equation}
	\begin{equation}
		F^{\mu\nu} = \partial^{\mu} A^{\nu} - \partial^{\nu} A^{\mu}.
	\end{equation}
	
	The original linear meson-exchange model, introduced by Walecka~~\cite{walecka1974}, provided a foundational framework but proved inadequate for quantitative nuclear predictions due to its overly stiff equation of state and insufficient flexibility in describing nuclear deformations~~\cite{gambhir1990,boguta1977}. To overcome these shortcomings, modern implementations introduce either nonlinear self-interaction terms or density-dependent coupling constants. For the nonlinear approach, a potential term is added to the Lagrangian:
	
	\begin{equation}
		U(\sigma) = \frac{1}{2} m_{\sigma}^2 \sigma^2 + \frac{1}{3} g_2 \sigma^3 + \frac{1}{4} g_3 \sigma^4,
	\end{equation}
	
	which has been successfully applied in various nuclear structure studies~~\cite{reinhard1986,lalazissis1997,toddrutel2005}. Alternatively, density-dependent couplings are introduced as:
	
	\begin{equation}
		g_i(\rho) = g_i(\rho_{\rm sat}) f_i(x), \quad i = \sigma, \omega, \rho,
	\end{equation}
	
	where $x = \rho / \rho_{\rm sat}$ is the ratio of local baryonic density to the saturation density of symmetric nuclear matter. In this case, nonlinear $\sigma$ terms are absent ($g_2 = g_3 = 0$). The functional form for the $\sigma$ and $\omega$ mesons is:
	
	\begin{equation}
		f_i(x) = a_i \frac{1 + b_i (x + d_i)^2}{1 + c_i (x + d_i)^2},
		\label{fx_me}
	\end{equation}
	
	while for the $\rho$ meson, it is:
	
	\begin{equation}
		f_{\rho}(x) = \exp(-a_{\rho} (x - 1)).
	\end{equation}
	
	These functions are constrained by $f_i(1) = 1$, $f_{\sigma}''(1) = f_{\omega}''(1)$, and $f_i''(0) = 0$, reducing the number of free parameters to three per meson type. In this work, we employ the DD-ME1~~\cite{niksic2002b} and DD-ME2~~\cite{lalazissis2005} parameterizations, with their values listed in Table~\ref{table:1}.
	
	\begin{table}[ht]
		\caption{Parameters of the DD-ME1~~\cite{niksic2002b} and DD-ME2~~\cite{lalazissis2005} parameterizations for the meson-exchange model.}
		\centering
		\begin{tabular}{c c c}
			\toprule
			Parameter & DD-ME1~~\cite{niksic2002b} & DD-ME2~~\cite{lalazissis2005} \\
			\hline
			$m$            & 939        & 939       \\
			$m_{\sigma}$   & 549.5255   & 550.1238  \\
			$m_{\omega}$   & 783        & 783       \\
			$m_{\rho}$     & 763        & 763       \\
			$g_{\sigma}$   & 10.4434    & 10.5396   \\
			$g_{\omega}$   & 12.8939    & 13.0189   \\
			$g_{\rho}$     & 3.8053     & 3.6836    \\
			$a_{\sigma}$   & 1.3854     & 1.3881    \\
			$b_{\sigma}$   & 0.9781     & 1.0943    \\
			$c_{\sigma}$   & 1.5342     & 1.7057    \\
			$d_{\sigma}$   & 0.4661     & 0.4421    \\
			$a_{\omega}$   & 1.3879     & 1.3892    \\
			$b_{\omega}$   & 0.8525     & 0.924     \\
			$c_{\omega}$   & 1.3566     & 1.462     \\
			$d_{\omega}$   & 0.4957     & 0.4775    \\
			$a_{\rho}$     & 0.5008     & 0.5647    \\
			\hline
		\end{tabular}
		\label{table:1}
	\end{table}
	
	\subsubsection{Point-Coupling Model}
	
	In contrast, the density-dependent point-coupling model employs zero-range interactions, represented by contact terms in the Lagrangian~~\cite{nikolaus1992, rusnak1997, burvenich2002}. The corresponding Lagrangian is:
	
	\begin{eqnarray}
		{\cal L} &=& \bar{\psi} (i \gamma \cdot \partial - m) \psi \nonumber \\
		&-& \frac{1}{2} \alpha_S(\hat{\rho}) (\bar{\psi} \psi) (\bar{\psi} \psi) - \frac{1}{2} \alpha_V(\hat{\rho}) (\bar{\psi} \gamma^{\mu} \psi) (\bar{\psi} \gamma_{\mu} \psi) \nonumber \\
		&-& \frac{1}{2} \alpha_{TV}(\hat{\rho}) (\bar{\psi} \vec{\tau} \gamma^{\mu} \psi) (\bar{\psi} \vec{\tau} \gamma_{\mu} \psi) \nonumber \\
		&-& \frac{1}{2} \delta_S (\partial_{\nu} \bar{\psi} \psi) (\partial^{\nu} \bar{\psi} \psi) - e \bar{\psi} \gamma \cdot A \frac{(1 - \tau_3)}{2} \psi.
		\label{lag_pc}
	\end{eqnarray}
	
	This formulation includes the free-nucleon term, isoscalar-scalar ($S$), isoscalar-vector ($V$), and isovector-vector ($TV$) interaction terms, a gradient term to capture finite-range effects, and the electromagnetic coupling for protons. The density dependence of the coupling strengths $\alpha_S$, $\alpha_V$, and $\alpha_{TV}$ is parameterized similarly to the meson-exchange model, tailored to nuclear matter properties. Here, we adopt the DD-PC1~~\cite{niksic2008} and DD-PCX~~\cite{yuksel2019} parameter sets, detailed in Table~\ref{table:2}.
	
	\begin{table}[ht]
		\caption{Parameters of the DD-PC1~~\cite{niksic2008} and DD-PCX~~\cite{yuksel2019} parameterizations for the point-coupling model.}
		\centering
		\begin{tabular}{c c c}
			\hline
			Parameter & DD-PC1~~\cite{niksic2008} & DD-PCX~~\cite{yuksel2019} \\
			\hline
			$m$ (MeV)      & 939         & 939         \\
			$a_s$ (fm$^2$) & -10.04616   & -10.97924   \\
			$b_s$ (fm$^2$) & -9.15042    & -9.03825    \\
			$c_s$ (fm$^2$) & -6.42729    & -5.31301    \\
			$d_s$          & 1.37235     & 1.37909     \\
			$a_v$ (fm$^2$) & 5.91946     & 6.43014     \\
			$b_v$ (fm$^2$) & 8.8637      & 8.87063     \\
			$d_v$          & 0.65835     & 0.65531     \\
			$b_{tv}$ (fm$^2$) & 1.83595  & 2.96321     \\
			$d_{tv}$       & 0.64025     & 1.30980     \\
			$\delta_s$ (fm$^4$) & -0.8149 & -0.87885    \\
			\hline
		\end{tabular}
		\label{table:2}
	\end{table}
	\subsection{Pairing Formalism}
	\label{subsec:pairing}

	Pairing correlations are essential for accurately describing open-shell nuclei,, particularly for neutron-rich isotopes approaching the neutron drip line~\cite{tanihata1996, Horowitz2001}. To account for both particle-hole and particle-particle channels in a unified framework, we employ the relativistic Hartree-Bogoliubov (RHB) model~\cite{Kucharek1991, ring1996}. In this approach, long-range mean-field effects are described through self-consistent meson fields, while short-range pairing correlations are incorporated via a pairing field $\hat{\Delta}$.
		
		The ground state within the RHB framework is expressed as a quasiparticle vacuum defined by the Bogoliubov transformation:
		\begin{equation}
			\alpha_k^\dagger = \sum_n U_{nk} c_n^\dagger + V_{nk} c_n,
		\end{equation}
		where $U_{nk}$ and $V_{nk}$ denote quasiparticle wave function components, and $n$ indexes both large ($f$) and small ($g$) components of the Dirac spinors.
		
		The total energy functional of the RHB method is written as:
		\begin{equation}
			E_{\text{RHB}}[\hat{\rho}, \hat{\kappa}] = E_{\text{RMF}}[\hat{\rho}] + E_{\text{pair}}[\hat{\kappa}],
		\end{equation}
		where $\hat{\rho}$ is the normal density matrix, and $\hat{\kappa}$ is the pairing tensor characterizing correlations between time-reversed states.
		
		The pairing contribution to the total energy is given by:
		\begin{equation}
			E_{\text{pair}}[\hat{\kappa}] = \frac{1}{4} \sum_{n_1 n_1^\prime} \sum_{n_2 n_2^\prime} \kappa^*_{n_1 n_1^\prime} \langle n_1 n_1^\prime | V^{pp} | n_2 n_2^\prime \rangle \kappa_{n_2 n_2^\prime},
		\end{equation}
		with $V^{pp}$ representing the effective interaction in the particle-particle channel.
		
		For this study, we adopt a finite-range separable pairing force, formulated in coordinate space. The interaction has the following structure:
		\begin{equation}
			V^{pp}(\boldsymbol{r}_1,\boldsymbol{r}_2,\boldsymbol{r}_1',\boldsymbol{r}_2') = 
			-G \, \delta(\boldsymbol{R} - \boldsymbol{R}') \, P(\boldsymbol{r}) \, P(\boldsymbol{r}'),
			\label{eq:pairing_force}
		\end{equation}
		where $\boldsymbol{R} = \frac{1}{\sqrt{2}}(\boldsymbol{r}_1 + \boldsymbol{r}_2)$ and $\boldsymbol{r} = \frac{1}{\sqrt{2}}(\boldsymbol{r}_1 - \boldsymbol{r}_2)$ are the center-of-mass and relative coordinates, respectively. The function $P(\boldsymbol{r})$ is a Gaussian of the form:
		\begin{equation}
			P(\boldsymbol{r}) = \frac{1}{(4\pi a^2)^{3/2}} \exp\left(-\frac{\boldsymbol{r}^2}{2a^2}\right),
			\label{eq:pairing_formfactor}
		\end{equation}
	
The parameters of the pairing interaction ($G = 728$ MeV$\cdot$fm$^3$ and $a = 0.644$ fm) were calibrated to replicate the pairing characteristics of the Gogny D1S force in symmetric nuclear matter~\cite{Berger1991}. This separable pairing interaction provides an excellent approximation to the original Gogny force while being computationally more efficient for large-scale calculations.

The pairing window in our calculations includes all positive-energy states up to 100 MeV above the Fermi surface, which ensures proper treatment of pairing correlations while avoiding ultraviolet divergence. For the solution of the RHB equations, we use a spherical box boundary condition with a radius of 20 fm and a step size of 0.1 fm, which provides an adequate treatment of continuum states in weakly-bound nuclei near the neutron dripline.

The numerical implementation employs a harmonic oscillator basis expansion with 20 oscillator shells for the calculation of matrix elements. The cutoff parameters for the pairing space were chosen to ensure convergence of the pairing energy to within 0.1 MeV. This pairing scheme has been extensively tested and provides a reliable description of pairing correlations across the nuclear chart \cite{Tian2009,TIAN2009}.

	\section{Results and Discussion}
	
	\subsection{Binding Energy}
	
\begin{figure*}[ht]
		\centering
		\includegraphics[scale=0.5]{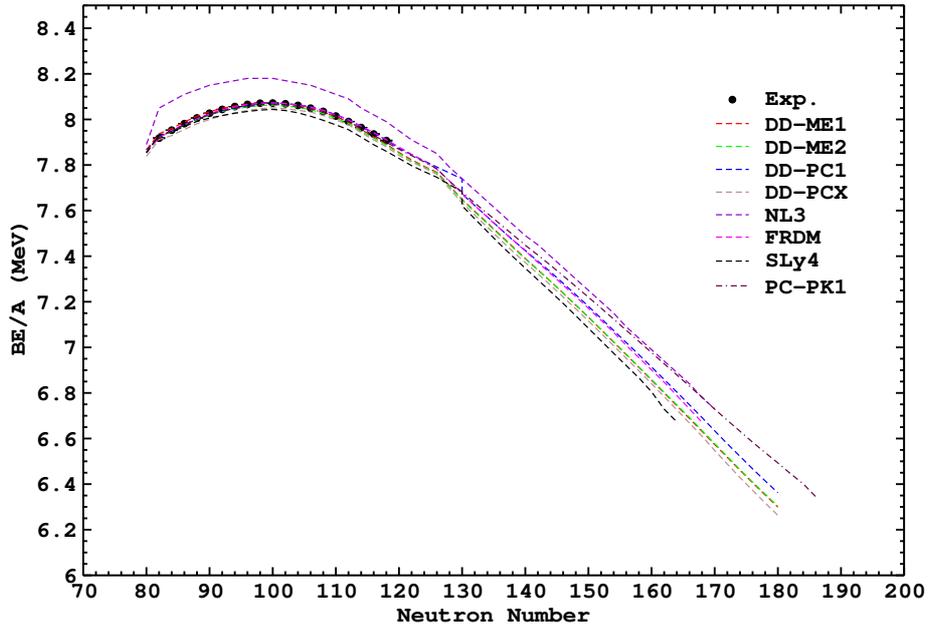}
		\caption{Binding energy per nucleon as a function of mass number for even-even hafnium isotopes ($^{152\text{--}270}$Hf), calculated using CDFT with DD-ME1, DD-ME2, DD-PC1, and DD-PCX interactions. Results are compared with predictions from RMF (NL3), DRHBc (PC-PK1), HFB (SLy4), FRDM, and experimental data~\cite{wang2021}.}
		\label{Hfbe}
	\end{figure*}

	The binding energy per nucleon serves as a key indicator of nuclear stability. Figure~\ref{Hfbe} presents the ground-state binding energy per nucleon for even-even hafnium isotopes, calculated within the CDFT framework using various density-dependent interactions: DD-ME1~\cite{niksic2002b}, DD-ME2~\cite{lalazissis2005}, DD-PC1~\cite{niksic2008}, and DD-PCX~\cite{yuksel2019}.
	
	Due to the scarcity of experimental data for neutron-rich nuclei near the drip line, it becomes essential to validate theoretical predictions through comparisons with alternative nuclear models~\cite{stoitsov2003,moller2016, Zhao2010, zhang2022, guo2024, mahapatro2015}. To this end, we have compared our results with those from other established theoretical frameworks, including the HFB + THO method using the Skyrme SLy4 interaction~\cite{stoitsov2003}, Finite Range Droplet Model (FRDM) predictions~\cite{moller2016},  the deformed relativistic Hartree–Bogoliubov theory in the continuum (DRHBc) utilizing the PC-PK1 functional~\cite{Zhao2010,zhang2022,guo2024}, and Relativistic Mean Field (RMF) calculations based on the NL3 interaction~\cite{mahapatro2015}. These comparisons reveal a high degree of agreement with our CDFT results, particularly in the neutron-rich regime approaching the drip line.
	
	The calculated binding energies show a coherent trend across the entire hafnium isotopic chain. Notably, the FRDM predictions closely match our results, even in the extreme neutron-rich region. The RMF (NL3) model tends to predict slightly more bound isotopes compared to our CDFT calculations, while the SLy4-based HFB results show a tendency toward lower binding energies.  Overall, the calculated results show a satisfactory level of agreement across both relativistic and non-relativistic models, as well as with the available experimental data~\cite{wang2021}.

	In general, the binding energy per nucleon (BE/A) increases with neutron number,  with neutron number and reaches a peak at  mass number $A = 172 (N=100)$ for all interactions considered. This behavior is consistent with experimental findings and indicates that $^{172}$Hf is the most stable isotope within the chain.
\

	\subsection{Quadrupole Deformation}
	
\begin{figure*}[ht]
		\centering
		\includegraphics[scale=0.5]{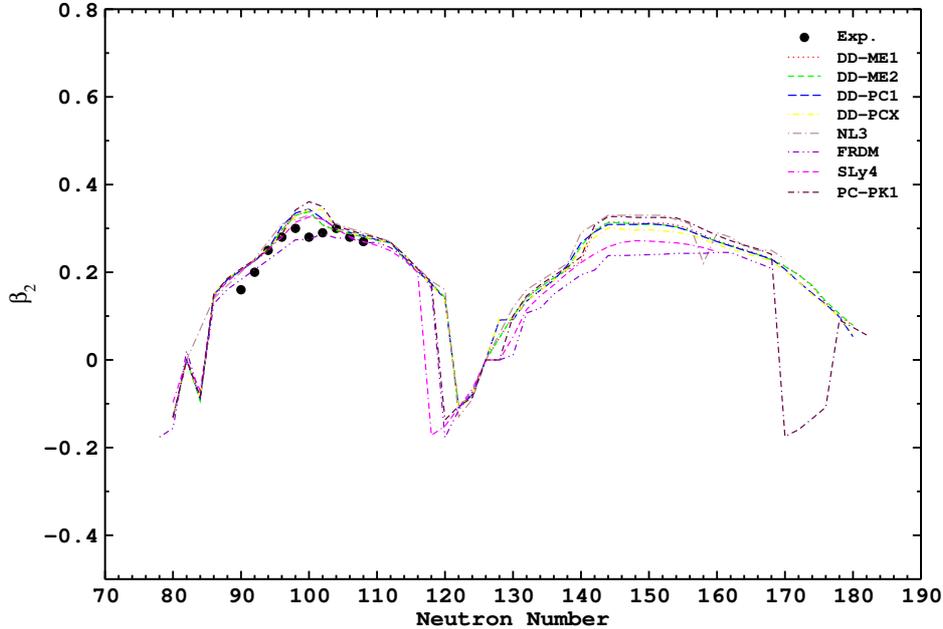}
		\caption{Quadrupole deformation parameter $\beta_2$ as a function of neutron number for even-even Hf isotopes. The CDFT results are shown for DD-ME1, DD-ME2, DD-PC1, and DD-PCX, and compared with RMF (NL3), DRHBc (PC-PK1), HFB (SLy4), FRDM predictions, and available experimental data~\cite{raman2001, pritychenko2016}.}
		
		\label{Hfbeta}
	\end{figure*}

	Nuclear shape and deformation are key physical parameters that significantly influence various nuclear properties, including nuclear size, isotopic shifts, and electric quadrupole moments. To illustrate the evolution of nuclear shapes along the hafnium isotopic chain, Fig.~\ref{Hfbeta} shows the variation of the quadrupole deformation parameter as a function of neutron number.
	
	To provide a more comprehensive analysis of shape transitions, the results obtained using the CDFT framework with density-dependent interactions are compared with predictions from other theoretical models. These include the RMF approach with the NL3 functional~\cite{mahapatro2015},  the DRHBc with the density functional PC-PK1~\cite{zhang2022,guo2024}, the non-relativistic HFB method with the Skyrme SLy4 interaction~\cite{stoitsov2003}, and FRDM~\cite{moller2016}. 	
	Additionally, available experimental data~\cite{raman2001, pritychenko2016} are used for comparison.
	
	Across the entire range of even-even hafnium isotopes, spanning from neutron-deficient to neutron-rich nuclei, we observe a sequence of shape transitions: from spherical to prolate, then prolate to oblate, followed by a return to spherical, and finally a transition back to prolate deformation. This complex pattern reflects the transitional character of hafnium isotopes.
	
	In the concerned isotopic region, our calculations using DD-ME1, DD-ME2, DD-PC1, and DD-PCX interactions show good agreement with the avaiable experimental data and with other theoretical predictions. However, slight deviations are observed in the neutron-rich region, particularly for  $N = 118$  to  $N = 122$ , when comparing our results with those from the SLy4-based HFB theory and FRDM.  	Notably, the Hf isotopes approaching the neutron drip line generally exhibit weak prolate deformation.  However, in this region, calculations based on the PC-PK1 functional within the DRHBc framework predict a slight oblate shape, in contrast to the predominantly prolate configurations suggested by other theoretical models.
	
	\subsection{Two-Neutron Separation Energy ($S_{2n}$), Shell Gap ($\delta S_{2n}$), and Neutron Pairing Energy ($E_{pair,n}$)}

	Identifying shell closures is crucial for a deeper understanding of nuclear structure. Two key observables commonly used to probe these closures are the two-neutron separation energy ($S_{2n}$) and the two-neutron shell gap ($\delta S_{2n}$).  The shell gap, also referred to as the differential of $S_{2n}$, is defined by the expression:
	
	$$\delta S_{2n}(N,Z) = \frac{S_{2n}(N,Z) - S_{2n}(N+2,Z)}{2}$$
	
\begin{figure*}[ht]
		\centering
		\includegraphics[scale=0.5]{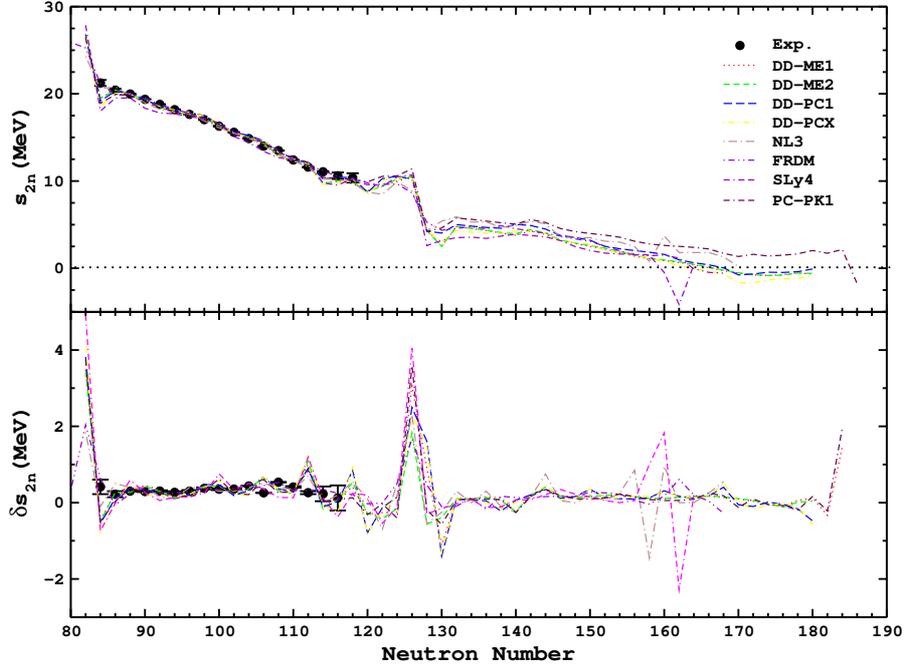}
		\caption{Two-neutron separation energy ($S_{2n}$, top panel) and two-neutron shell gap ($\delta S_{2n}$, bottom panel) as functions of neutron number for even-even Hf isotopes calculated using CDFT with four interactions. Results are compared with experimental data, RMF (NL3), DRHBc (PC-PK1), HFB (SLy4), and FRDM models.}
		
		\label{Hfsn}
	\end{figure*}

	In Fig.~\ref{Hfsn}, we present the calculated values of $S_{2n}$ and $\delta S_{2n}$ for even-even $^{152\text{--}270}$Hf isotopes. Our calculations are performed within the CDFT framework using several density-dependent effective interactions (DD-ME1, DD-ME2, DD-PC1, and DD-PCX). To assess the reliability of our results, we compare them with available experimental data as well as with theoretical predictions from other models, including the RMF approach with the NL3 functional~\cite{mahapatro2015}, the DRHBc approach with the density functional PC-PK1~\cite{zhang2022,guo2024}, the non-relativistic HFB method with the Skyrme SLy4 interaction~\cite{stoitsov2003}, and the Finite Range Droplet Model (FRDM)~\cite{moller2016}.
	Notably, the Hf isotopes approaching the neutron drip line generally exhibit weak prolate deformation. However, in this region, calculations based on the PC-PK1~\cite{zhang2022,guo2024} functional within the DRHBc framework predict a slight oblate shape, in contrast to the predominantly prolate configurations suggested by other theoretical models.
	
	The location of the neutron drip line represents a fundamental limit of nuclear existence and provides crucial insights into the stability of neutron-rich nuclei. From our calculations of the two-neutron separation energy ($S_{2n}$), it is evident that the neutron drip line for even-even hafnium isotopes occurs around neutron number $N \approx 170$ for all the CDFT parameterizations considered (DD-ME1, DD-ME2, DD-PC1, and DD-PCX). Beyond this point, $S_{2n}$ becomes negative, indicating that additional neutrons cannot be bound to the nucleus.
	
	This prediction is in good agreement with other theoretical approaches. In particular, the RMF model with the NL3 interaction~\cite{mahapatro2015} and the FRDM~\cite{moller2016} also predict the neutron drip line to lie near $N \approx 170$ for hafnium isotopes. 
	{In contrast, the HFB approach employing the non-relativistic Skyrme SLy4 force~\cite{stoitsov2003} predicts an earlier termination of bound isotopes, around $N \approx 160$, indicating a narrower range of neutron-rich nuclei within that model. On the other hand, calculations using the DRHBc model with the PC-PK1 functional~\cite{zhang2022,guo2024} extend the drip line further to $N = 184$, highlighting the sensitivity of drip line predictions to the underlying theoretical framework and interaction.

	These variations highlight the sensitivity of drip line predictions to the choice of interaction and underline the importance of using multiple theoretical models when exploring the limits of nuclear stability.
	
	As shown in Fig.~\ref{Hfsn}, we observe sudden drops in $S_{2n}$ and prominent peaks in $\delta S_{2n}$ at neutron numbers $N = 82$ and  $N = 126$ , clearly indicating the presence of established neutron shell closures. However, in this region, calculations based on the PC-PK1 functional within the DRHBc framework predict a slight oblate shape, in contrast to the predominantly prolate configurations suggested by other theoretical models. Interestingly, an additional kink appears around $ N = 118$  in our CDFT results (across all four interactions), which is also supported by predictions from the RMF (NL3)~\cite{mahapatro2015}, the DRHBc (PC-PK1)~\cite{zhang2022,guo2024}, and HFB (SLy4)~\cite{stoitsov2003} models, suggesting a possible subshell closure at this neutron number. In contrast, the FRDM model does not exhibit any notable signature at $N = 118 $.
	These findings are consistent with earlier studies proposing the existence of deformed subshell closures~\cite{karim2015, karim2018, naz2019}.

	Further support for shell closures comes from the behavior of the neutron pairing energy ($E_{\text{pair},n}$), which typically vanishes at magic numbers~\cite{delestal2001, sil2004}. As shown in Fig.~\ref{Hfep}, $E_{\text{pair},n}$ drops to zero at  $N = 82$ and $ N = 126$ , reinforcing the shell gaps observed in $S_{2n}$ and $\delta S_{2n}$. Interestingly, the DD-PCX interaction predicts a vanishing pairing energy at $ N = 118$, while other functionals (DD-ME1, DD-ME2, and DD-PC1) show near-zero values, consistent with the subshell behavior suggested in Fig.~\ref{Hfsn}.
	
	These results provide valuable insights into the shell evolution and structural changes in hafnium isotopes, particularly in the neutron-rich regime. The possible presence of a subshell closure at  $N = 118 $ highlights the intricate nature of shell structure in heavy nuclei and calls for further experimental and theoretical investigations.
	
	\begin{figure*}[ht]
		\centering
		\includegraphics[scale=0.5]{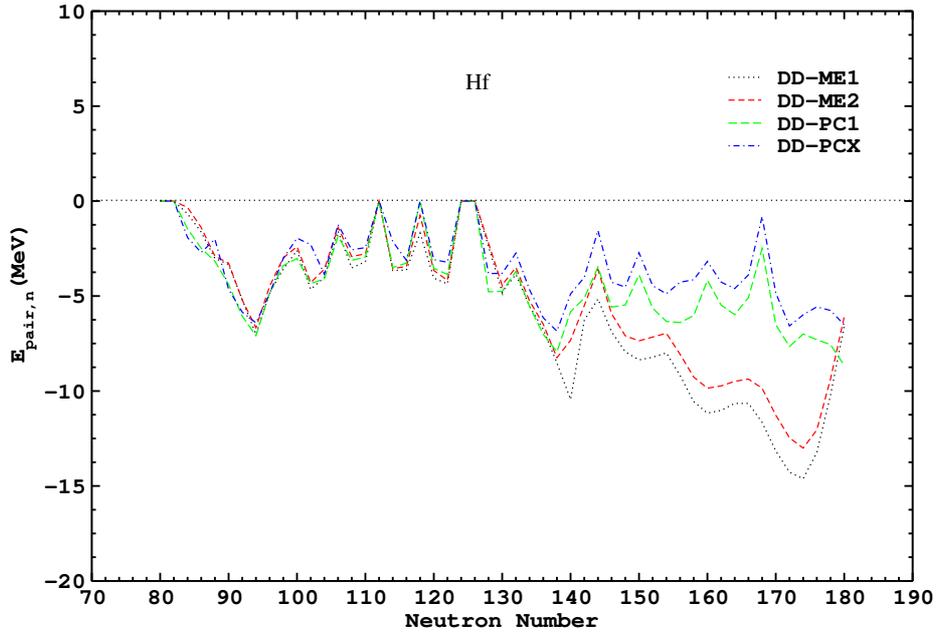}
		\caption{Neutron pairing energy ($E_{\text{pair},n}$) as a function of neutron number for even-even Hf isotopes calculated with CDFT (DD-ME1, DD-ME2, DD-PC1, DD-PCX). Vanishing or near-zero values at $N = 82$, $N = 126$, and $N = 118$ indicate possible shell or subshell closures.}
		\label{Hfep}
	\end{figure*}

	\subsection{Neutron, Proton, and Charge Radii}
	
	\begin{figure*}[ht]
		\centering
		\includegraphics[scale=0.5]{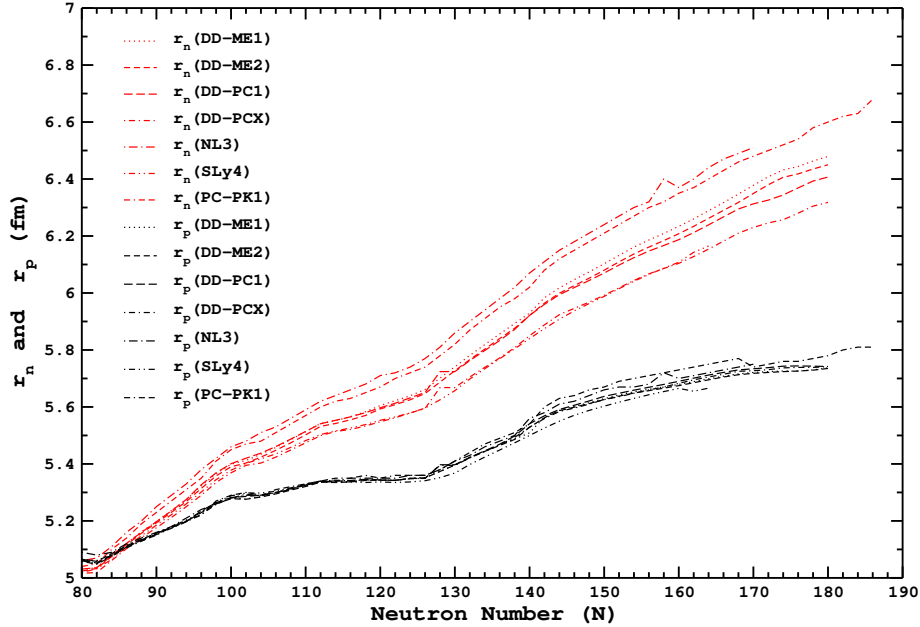}
		\caption{RMS neutron radius ($r_n$) and proton radius ($r_p$) for even-even Hf isotopes as calculated with CDFT using four density-dependent interactions. Results are compared with RMF (NL3), DRHBc (PC-PK1), and HFB (SLy4) models. Discontinuities at $N = 82$ and $N = 126$ reflect closed-shell behavior.}
		
		\label{Hfrad}
	\end{figure*}

	Nuclear radii offer vital insights into the size and spatial distribution of protons and neutrons within atomic nuclei. Among these, the neutron skin thickness, defined as the difference between the neutron and proton root-mean-square (RMS) radii, plays a key role in characterizing neutron-rich nuclei. This quantity becomes particularly important when there is a significant neutron excess and has deep connections to the isospin dependence of the nuclear equation of state, thereby linking finite nuclei with the properties of infinite nuclear matter~\cite{reinhard2016,hagen2016,agrawal2006}.
	
	Figure~\ref{Hfrad} presents the calculated evolution of neutron ($r_n$) and proton ($r_p$) RMS radii along the even-even Hf isotopic chain. A gradual increase in $r_n$ is observed as additional neutrons are added, consistent across all interactions. However, discontinuities at neutron numbers $N = 82$ and $N = 126$ clearly signal shell closures, with corresponding dips in the growth of neutron radii. Proton radii ($r_p$), in contrast, remain relatively stable across the chain due to the fixed proton number ($Z = 72$).
	
	Shape transitions and changes in deformation—particularly in mid-shell regions—introduce subtle variations in radii, which are model-dependent. For instance, predictions from the DRHBc model using PC-PK1 tend to yield slightly smaller $r_p$ values compared to CDFT models, while HFB (SLy4) results show more compressed neutron radii for heavier isotopes. RMF (NL3) consistently predicts slightly larger neutron radii than other frameworks, suggesting a stiffer isovector interaction.

	\begin{figure*}[ht]
		\centering
		\includegraphics[scale=0.5]{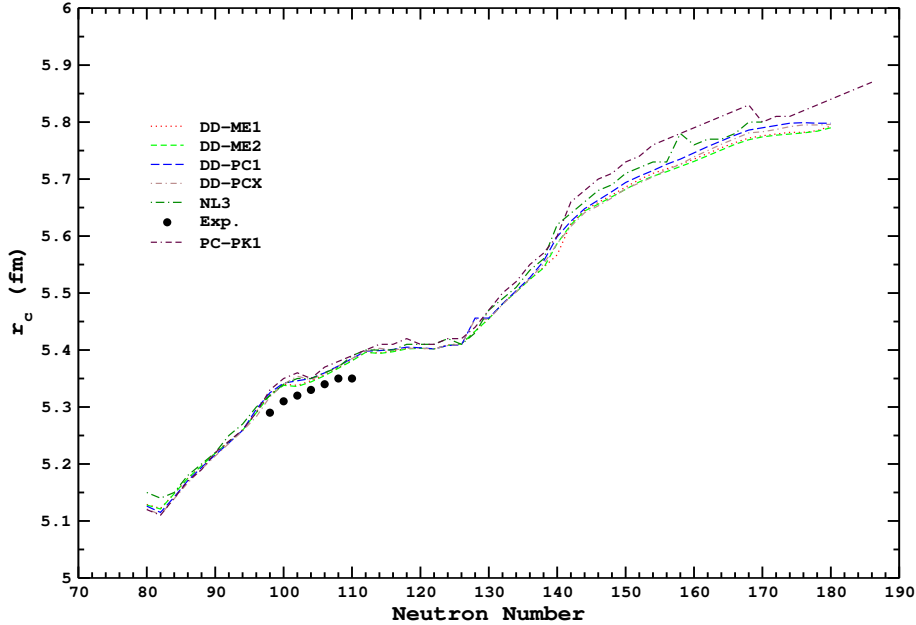}
		\caption{Root-mean-square (RMS) charge radii ($r_c$) of even-even Hf isotopes obtained from CDFT calculations using DD-ME1, DD-ME2, DD-PC1, and DD-PCX interactions. Theoretical results are compared with available experimental data~\cite{angeli2004, angeli2013}, RMF (NL3), and DRHBc (PC-PK1) models.}
		\label{Hfrc}
	\end{figure*}
	
   The RMS charge radii ($r_c$), displayed in Fig.~\ref{Hfrc}, show good agreement between our CDFT results and available experimental measurements~\cite{angeli2004,angeli2013}. All four density-dependent CDFT interactions (DD-ME1, DD-ME2, DD-PC1, and DD-PCX) accurately capture the overall trend. Both RMF (NL3) and DRHBc (PC-PK1) tend to slightly overestimate the charge radii of heavier Hf isotopes, particularly beyond $N > 140$. This discrepancy may arise from differences in the treatment of deformation effects and pairing correlations within these models.
	
	\begin{figure*}[ht]
		\centering
		\includegraphics[scale=0.5]{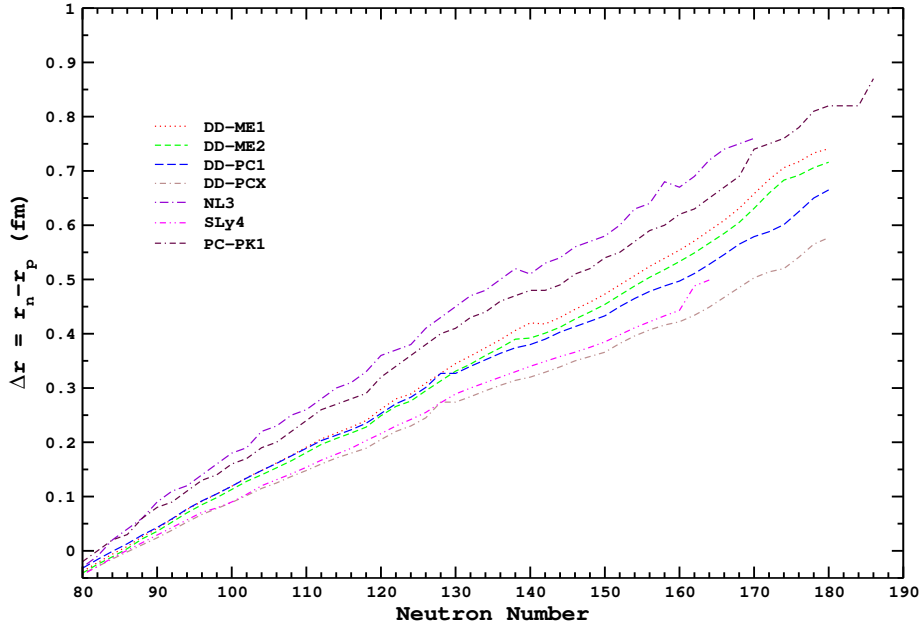}
		\caption{Neutron skin thickness ($\Delta r$) for even-even Hf isotopes as a function of neutron number, showing monotonic increase with neutron excess. Results based on CDFT calculations using DD-ME1, DD-ME2, DD-PC1, and DD-PCX interactions are compared with RMF (NL3), DRHBc (PC-PK1), and HFB (SLy4).}
		\label{Hfskin}
	\end{figure*}
	
	As seen in Fig.~\ref{Hfskin}, the neutron skin thickness $\Delta r$ exhibits a steady rise with increasing neutron number. This trend confirms the expected enhancement of the neutron-rich surface layer in heavier isotopes and highlights the connection between isospin asymmetry and nuclear structure.

	\subsection{Potential Energy Curve}
	
	The nuclear shape, characterized by quadrupole deformation, is a fundamental feature that significantly influences various nuclear properties. Most nuclei adopt either spherical or ellipsoidal shapes. In the axially symmetric approximation, the degree of deformation is quantified by the quadrupole deformation parameter $\beta_2$. In this work, we employ Covariant Density Functional Theory (CDFT) using density functionals such as DD-ME1~\cite{niksic2002b}, DD-ME2~\cite{lalazissis2005}, DD-PC1~\cite{niksic2008}, and DD-PCX~\cite{yuksel2019} to carry out quadrupole-constrained calculations and obtain the potential energy curves (PECs) for even-even hafnium isotopes. 
	The resulting PECs are displayed in Figs.~\ref{ps1}–\ref{ps3}.
	
	\begin{figure*}[hbt!]
		\centering
		\includegraphics[scale=0.5]{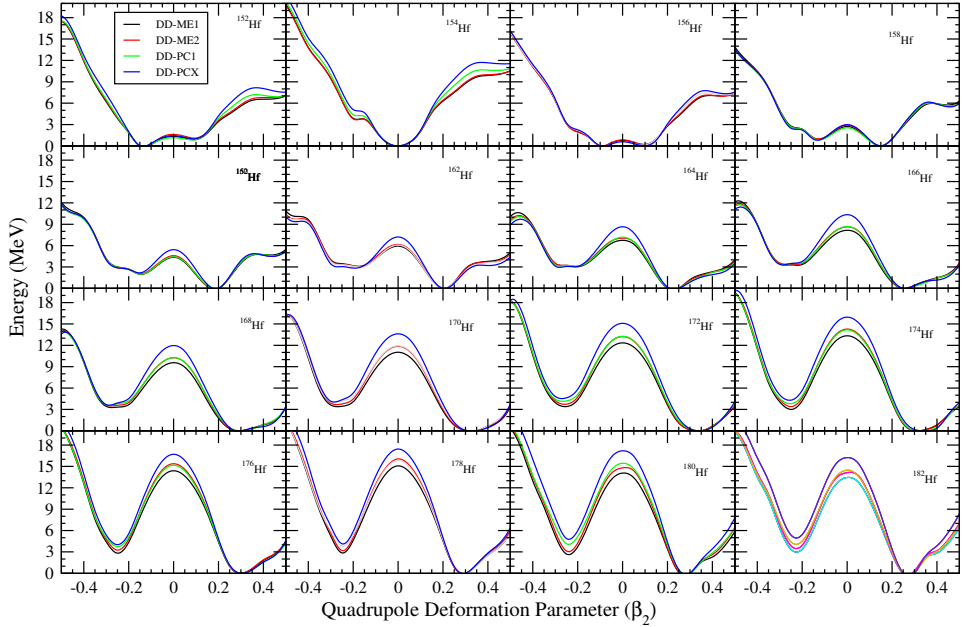}
		\caption{Potential energy curves for the even-even isotopes $^{152\text{--}182}$Hf as functions of the quadrupole deformation parameter $\beta_2$ for DD-ME1, DD-ME2, DD-PC1, and DD-PCX interactions. Energies are normalized to the ground state.}
		\label{ps1}
	\end{figure*}
	
	\begin{figure*}[hbt!]
		\centering
		\includegraphics[scale=0.5]{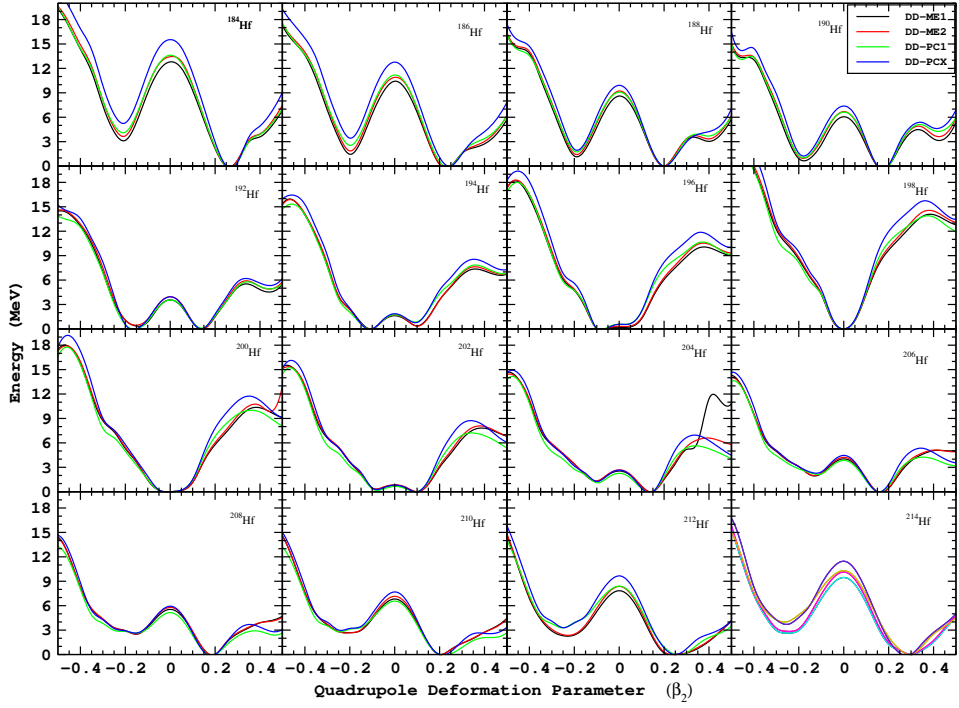}
		\caption{Same as Fig.~\ref{ps1}, but for $^{184\text{--}214}$Hf isotopes.}
		\label{ps2}
	\end{figure*}
	
	\begin{figure*}[hbt!]
		\centering
		\includegraphics[scale=0.5]{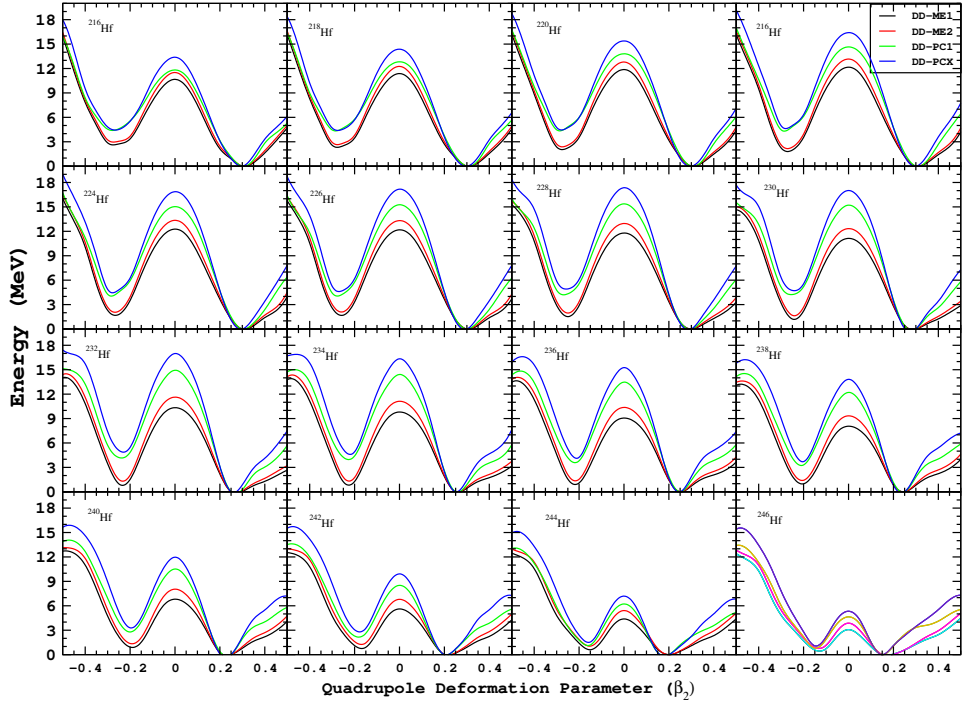}
		\caption{Same as Fig.~\ref{ps1}, but for $^{216\text{--}246}$Hf isotopes.}
		\label{ps3}
	\end{figure*}
	
	The results reveal minimal differences between the meson-exchange and point-coupling interactions, with the positions of the spherical, prolate, and oblate minima remaining consistent across models. However, slight variations in the relative energies of these minima can be observed, depending on the interaction used. Overall, the energy barriers and deformation behavior are found to be broadly consistent among the different functionals.
	
	For instance, the ground state of $^{154}$Hf is predicted to be spherical, consistent with the well-known neutron shell closure at $ N = 82$ . As the neutron number increases, the energy minimum shifts toward the prolate side. The isotopes $^{156}$Hf and $^{158}$Hf exhibit shape coexistence, with nearly degenerate prolate and oblate configurations differing by less than $1.0$ MeV. For $^{160\text{--}190}$Hf, an oblate excited minimum is observed $1.2-2.5$ MeV above the global prolate minimum. Notably, Fig.~\ref{ps2} illustrates strong competition between multiple low-lying configurations of differing intrinsic shapes, indicative of shape coexistence.
	
	In $^{192}$Hf and $^{194}$Hf, both prolate and oblate minima are found at $\beta_2 = \pm 0.15$, with energy differences of approximately 0.8 MeV and 0.3 MeV, respectively. Around $^{198}$Hf, a clear transition from deformed to spherical shapes is observed, with a sharp minimum at $\beta_2 = 0.0$, attributed to the neutron shell closure at \( N = 126 \).
	
	Beyond this point, in the neutron-rich region, spherical symmetry is lost again, and well-deformed prolate ground states reemerge. Isotopes in the range $^{202\text{--}220}$Hf exhibit strong prolate deformation, with prolate minima lying $2.0-3.0$ MeV below the oblate ones in DD-ME1 and DD-ME2, and $2.0-5.0$ MeV in DD-PC1 and DD-PCX calculations. Similarly, $^{222\text{--}240}$Hf isotopes maintain prolate shapes, with prolate–oblate energy differences ranging from $1.5-2.5$ MeV (DD-ME1, DD-PC1) and $2.0-4.0$ MeV (DD-ME2, DD-PCX). Interestingly, $^{232}$Hf and $^{236}$Hf show near-degenerate prolate and oblate minima in DD-PC1, suggesting possible shape coexistence, a feature not observed in other interactions.
	
	Our findings are in qualitative agreement with earlier studies reported in Refs.~\cite{Sarriguren2008, hartley2005, vasileiou2023, afana2016, choi2022}, and further support the presence of shape transitions and coexistence phenomena in hafnium isotopes.

	\section{Conclusion}
	
	In this study, we have performed a comprehensive study of even-even hafnium isotopes using the framework of CDFT, employing four modern density-dependent interactions: DD-ME1, DD-ME2, DD-PC1, and DD-PCX. A range of ground-state and collective properties were systematically analyzed, including binding energy, quadrupole deformation, two-neutron separation energy ($S_{2n}$), shell gap ($\delta S_{2n}$), neutron pairing energy ($E_{\text{pair},n}$), nuclear radii, neutron skin thickness, and potential energy curves (PECs).
	
	The calculated binding energies per nucleon demonstrate excellent agreement with available experimental data and other established theoretical models such as FRDM, RMF (NL3), and HFB (SLy4). Among the isotopic chain, $^{172}$Hf emerges as the most bound isotope, indicating its enhanced stability.
	
	Quadrupole deformation analysis reveals a rich variety of shape transitions along the hafnium isotopic chain, encompassing spherical, prolate, and oblate configurations. Our results confirm the transitional nature of Hf nuclei, with noticeable shape coexistence in select isotopes, especially in the mid-shell regions. Deviations observed in neutron-rich regions, particularly near $N=118$, reflect the sensitivity of deformation behavior to the underlying nuclear interaction. Interestingly, DRHBc calculations using the PC-PK1 functional indicate weak oblate deformation in some neutron-rich isotopes, contrasting with the predominantly prolate shapes predicted by other models.
	
	A detailed study of the two-neutron separation energy and shell gap highlights pronounced shell closures at $N=82$ and $N=126$, as evidenced by sharp drops in $S_{2n}$ and peaks in $\delta S_{2n}$. Importantly, the two-neutron separation energy systematics predict the neutron drip line to occur around $N \approx 170$ in all CDFT interactions, consistent with RMF (NL3) and FRDM predictions, while SLy4 indicates an earlier termination near $N \approx 160$. In contrast, the DRHBc framework with PC-PK1 extends the drip line prediction further to $N=184$. Additionally, our calculations indicate a possible subshell closure at $N=118$, supported by multiple interactions within CDFT as well as other theoretical approaches like RMF and HFB. This observation is further reinforced by the vanishing neutron pairing energy around $N=118$, suggesting a weakened pairing correlation due to the emergence of a subshell gap.
	
	The nuclear radius and neutron skin thickness analyses provide valuable insights into matter distribution in neutron-rich isotopes. A gradual increase in neutron skin thickness with neutron number was observed, consistent with the growing neutron excess. Theoretical predictions for charge radii show satisfactory agreement with experimental measurements, and the dips in neutron radius at magic numbers reinforce shell closure signatures. Notably, RMF (NL3) and DRHBc (PC-PK1) slightly overestimate the charge radii for heavier isotopes, especially beyond $N>140$, which may stem from differences in their treatment of deformation and pairing correlations. Although the present work focuses on four density-dependent CDFT interactions, we acknowledge that recently developed functionals such as PCF-PK1~\cite{Zhao2022} offer promising predictive power. Incorporating PCF-PK1 into future large-scale calculations could provide further insight into structure phenomena near the drip line.

	The potential energy curves (PECs) across the Hf isotopic chain exhibit diverse deformation landscapes, including spherical, prolate, oblate, and shape-coexistent configurations. The results show strong shape transitions, particularly around $N=82$ and $N=126$, in agreement with known shell effects. Notably, the reappearance of well-deformed prolate ground states in the heavy neutron-rich region and the presence of shape coexistence in isotopes like $^{158}$Hf and $^{232}$Hf point to complex structural behavior in this mass region.
	
	In summary, our findings provide a unified and consistent description of the structure and shape evolution of hafnium isotopes. The agreement with experimental data and established models validates the reliability of the CDFT approach with modern density-dependent interactions. The identification of a possible subshell closure at $N=118$ and regions of shape coexistence offer promising directions for future theoretical and experimental studies in heavy nuclei, particularly those approaching the neutron drip line.

	\section*{Acknowledgements}
	Usuf Rahaman would like to acknowledge the Department of Physics, Aligarh Muslim University for computational support.


\end{document}